\documentstyle[epsf,sprocl]{article}
\newcommand{\beq}    {\begin{equation}}
\newcommand{\eeq}    {\end{equation}}
\newcommand{\beqarr} {\begin{eqnarray}}
\newcommand{\eeqarr} {\end{eqnarray}}
\newcommand{\barr}   {\begin{array}}
\newcommand{\earr}   {\end{array}}
\newcommand{\no}     {\nonumber}

\newcommand{\lsim}{\mathrel{\mathop{\kern 0pt \rlap
  {\raise.2ex\hbox{$<$}}}
  \lower.9ex\hbox{\kern-.190em $\sim$}}}
\newcommand{\gsim}{\mathrel{\mathop{\kern 0pt \rlap
  {\raise.2ex\hbox{$>$}}}
  \lower.9ex\hbox{\kern-.190em $\sim$}}}

\newcommand{\mb}[1]  {\mbox{#1}}

\begin{document}

\title {NEUTRALINO DARK MATTER}

\author{ A. Bottino \footnote{Invited talk at the International Workshop on the 
"Identification of Dark Matter", Sheffield, UK, September 1996}}
\address{
Dipartimento di Fisica Teorica, Universit\`a di Torino\\
and\\
INFN, Sezione di Torino,
Via P. Giuria 1, 10125 Torino, Italy
}

\maketitle
\begin{abstract}
  We examine the main properties of relic neutralinos in the 
context of the Minimal Supersymmetric extension of the Standard Model.
We present a whole set of results obtained for the relic abundance and for 
detection rates with a model  constrained  according to the 
 latest accelerator data. Predicted detection rates for relic 
neutralinos are compared to the present experimental bounds. 
A comparison with results obtained  with  more 
sophisticated theoretical schemes with unification assumptions at 
$M_{GUT}$ is presented too.
We also briefly discuss some properties of a 
Higgsino-like 
neutralino which has recently been suggested in connection with a 
supersymmetric interpretation of a single event at the Fermilab Tevatron.

\end{abstract}

\section{Introduction}

  The neutralino is currently being considered as a favourite 
candidate for Cold Dark Matter (CDM). In fact,  the neutralino 
 turns out to be the Lightest Supersymmetric Particle (LSP)
in extended regions of the supersymmetric parameter space and then, 
if protected by R-parity conservation (that we assume to hold), it would 
have decoupled from the
primeval plasma and would be present in the Universe as a fossil particle. 

The properties of the neutralino vary quite noticeably as one moves in the
supersymmetric parameter space: from some regions where the neutralino relic
abundance $\Omega_{\chi} h^2$ ($h$ is the present Hubble expansion rate 
in units
of $100 ~\mbox{Km} \cdot \mbox{s}^{-1} \cdot \mbox{Mpc}^{-1}$)
     turns out to be large (and detection rates low) to other 
regions where the other extreme situation of a negligible 
$\Omega_{\chi}$ (and of sizable detection rates) occurs. 
It is remarkable that there are domains of the supersymmetric parameter space, 
where the neutralino can  provide a relic abundance in the preferred
density range for CDM \cite{ours}

\begin{equation}
\Omega_{CDM}  h^2 = 0.2 \pm 0.1.
\label{eq:nrange}
\end{equation}

However, even in the case when the neutralino does not contribute 
significantly to the average density of the Universe, experimental searches 
for relic neutralinos would be of the utmost interest for a number of 
 cosmological and particle-physics aspects.

  In this paper we examine the main properties of relic neutralinos in the 
context of the Minimal Supersymmetric extension of the Standard Model 
(MSSM)\cite{Susy}.
Indeed, this scheme  provides a most convenient phenomenological 
framework at the electroweak scale ($M_Z$), without assuming too 
strong, arbitrary theoretical hypotheses. 
After a short presentation of the theoretical scheme, we report 
the main results concerning neutralino relic abundance and predicted 
detection rates. A comparison with results obtained  with  more 
sophisticated theoretical schemes is presented too.

Furthermore, we also briefly discuss some properties of a 
Higgsino-like 
neutralino which has recently been suggested in connection with a 
supersymmetric interpretation of a single event at the Fermilab Tevatron. 

\section{The Minimal Supersymmetric Standard Model}

The MSSM is based on  the same gauge group as the Standard Model
and contains the supersymmetric extension of its particle content. The
Higgs sector is modified as compared to that of the Standard
Model, because it requires
two Higgs doublets $H_1$ and $H_2$ in order to give mass both to down-- and
up--type quarks and to cancel anomalies. After Electro--Weak Symmetry
Breaking (EWSB),
the physical Higgs fields consist of two
charged particles and three neutral ones: two scalar fields ($h$ and $H$) and
one pseudoscalar ($A$). The Higgs sector is specified at the tree level by
two independent parameters:
the mass of one of the physical Higgs fields (we will use the mass $m_A$ of the 
$A$ boson) and the ratio of the two vacuum
expectation values, usually defined as $\tan\beta=v_2/v_1 \equiv
<H_2> / <H_1>$.

Apart from the Yang-Mills Lagrangian, other characteristic elements of the 
model are the superpotential, 
which contains  all the Yukawa interactions
and the Higgs-mixing term 
$\mu H_1 H_2$, and  the soft--breaking
Lagrangian, which models the breaking of  supersymmetry 
\begin{eqnarray}
-{\cal L}_{soft}&=&
\displaystyle \sum_i m_i^2 |\phi_i|^2 +  \left\{\left[
A^{l}_{ab} h_{ab}^{l} \tilde{L_a} H_1 \tilde{R_b} +
A^{d}_{ab} h_{ab}^{d} \tilde{Q_a} H_1 \tilde{D_b}  \right. \right. \no \\ 
&+& \left.  \left. A^{u}_{ab} h_{ab}^{u} 
\tilde{Q_a} H_2 \tilde{U_b} +\mb{h.c.} \right]
-  \displaystyle  B \mu H_1 H_2 + \mb{h.c.}  \right\} \no \\
&+&  \sum_i M_i
(\lambda_i \lambda_i + \bar\lambda_i \bar\lambda_i).
\label{eq:soft}
\end{eqnarray}
\noindent
Here the notations are the following: 
 the $\phi_i$ are  the scalar fields, the $\lambda_i$ are the
gaugino fields, 
$\tilde Q$ and $\tilde L$
are the doublet squark and slepton fields, respectively,
and $\tilde U$, $\tilde D$ and
$\tilde R$ denote the $SU(2)$--singlet fields for the up--squarks,
down--squarks and sleptons, $m_i$ and $M_i$ are the mass
parameters of the scalar and gaugino fields, respectively. $A$ 
denotes the trilinear and $B$ the bilinear supersymmetry breaking parameters. 
The
Yukawa interactions are described by the parameters $h$, which
are related to the masses of the standard fermions by the usual
expressions, {\em e.g.}, $m_t = h^t v_2$.

The tree--level scalar potential for the neutral Higgs fields
turns out to be 
\begin{equation}
V_0 = (M_{H_1}^2+\mu^2) |H_1|^2 + (M_{H_2}^2+\mu^2) |H_2|^2 -B\mu (H_1 H_2 +
\mb{h.c.}) + \mb{quartic D terms.}
\label{eq:higgspot}
\end{equation}
\noindent
When the electro-weak symmetry breaking is triggered by $H_1$ and $H_2$
acquiring vacuum expectation values $v_1$, $v_2$ with the condition 
$v_1^2 + v_2^2 \equiv v^2 = 2 m_W^2 / g^2$, 
the potential $V_0$ takes the standard
form of a Higgs potential, provided the following contraints are satisfied:

\begin{equation}
\sin 2 \beta = \frac {-2B\mu} {M_{H_1}^2+M_{H_2}^2+2 \mu^2}
\label{eq:s2beta}
\end{equation}
\begin{equation}
M_Z^2 = 2 \frac {M_{H_1}^2-M_{H_2}^2 \tan^2 \beta}
{\tan^2 \beta -1} - 2 \mu^2.
\label{eq:mz}
\end{equation}
\noindent 
Notice that the sign of $\mu$ is defined according to the convention of
Refs. \cite{Susy}.
We  remark that 
Eqs.(\ref{eq:higgspot}--\ref{eq:mz}) are expressed
at the tree level. However, in the 
calculations that we will report later, 1--loop corrections
to $V_0$ are included \cite{oneloop}.

As is manifest from the previous formulae, the theoretical model 
contains an exceedingly large number of parameters. In order to have a
viable phenomenological model we reduce the number of free parameters by
introducing the following usual assumptions:

\begin{itemize}

\item $A_{33}^u = A_{33}^d = A_{33}^l \equiv A$; all other $A_{ab}$'s are set 
to zero

\item all squarks are taken as degenerate: $m_{\tilde q_i} \equiv m_0$, except 
for $m_{\tilde t}$

\item  all sleptons  are taken as degenerate with $m_{\tilde l_i} \equiv m_0$ 

\item the gaugino masses are assumed to be renormalised as the gauge couplings,

$M_1 : M_2 : M_3 = \alpha_1 : \alpha_2 : \alpha_3$; 
thus in particular $M_1= (5/3) \tan^2 \theta_W M_2$. 

\end{itemize}

We emphasize that all these relations are meant to be taken at the $M_Z$ scale.

It is worth noticing that any variation in some of these assumptions, 
if required, can be introduced staightforwardly without implying any other 
modification in the general scheme. The flexibility of this phenomenological 
approach is not shared by other more theoretically-constrained models.

In force of the  conditions previously discussed, the number of independent 
parameters is reduced to seven. We choose them to be: $M_2, \mu, \tan \beta, 
m_A, m_0, m_{\tilde t},\\ A$. 
In terms of these parameters, 
the supersymmetric space is further constrainted by
the following requirements: 

\begin{itemize}

\item all experimental bounds on Higgs, neutralino, chargino and
sfermion masses  are satisfied (taking into account also the new data from
LEP2 \cite{LEP2}) 

\item the neutralino is the Lightest Supersymmetric
Particle (LSP) (i.e., regions where gluino or squarks or sleptons are LSP are 
excluded) 

\item  constraints due to the $b \rightarrow s + \gamma$ process are taken into
account \cite{alam}

\item  the neutralino relic
abundance does not exceed the cosmological bound, i.e. 
$\Omega_{\chi}h^2 \leq 1$.

\end{itemize}

As usual, any neutralino mass--eigenstate is written as a linear superposition 

\begin{equation}
\chi_i = a_i \tilde \gamma + b_i \tilde Z + c_i \tilde {H_s} 
+d_i \tilde {H_a}
\label{eq:neu}
\end{equation}

\noindent
where $\tilde \gamma, \tilde Z$ are the photino and zino states and 
$\tilde {H_s}, \tilde {H_a}$ are defined by 
$\tilde {H_s} = {\rm sin} \beta \tilde H_1^{\circ} + 
{\rm cos} \beta \tilde H_2^{\circ}$, 
$\tilde {H_a} = {\rm cos} \beta \tilde H_1^{\circ} 
- {\rm sin} \beta \tilde H_2^{\circ}$, in terms of the higgsino fields 
$\tilde H_1^{\circ}$, $\tilde H_2^{\circ}$, supersymmetric partners of the Higgs
fields $H_1^{\circ}$, $H_2^{\circ}$. 
In the following we will simply call neutralino  the eigenstate of 
lowest-mass and use $\chi \equiv \chi_1$. Fig.\ref{fig1} gives the isomass 
contours in the $\mu - M_2$ plane. 
 
\begin{figure}[htb]
\epsfxsize=10cm
\vspace{14cm}
\caption{Isomass and isocomposition contour lines for the neutralino 
in the $\mu-M_2$ plane for $\tan \beta = 10$. The region excluded by LEP 
is denoted by dots. $m_{\chi}$ is in units of GeV. \label{fig1}}
\end{figure}
  
In the same plot, the region excluded by the LEP data and 
the contour 
lines for the composition parameter $P \equiv a_1^2 + b_1^2$ are also 
displayed.

Various properties of relic neutralinos in the MSSM have been considered by 
a number of authors. Some of the most recent ones are given in 
Refs.\cite{jkg,bg}. 
In this review,  we present a whole  set of results 
concerning the main properties for relic 
neutralinos (relic abundance and detection rates) 
performed in the same consistent theoretical framework 
\cite{all}.
The techniques employed for the calculation of these various 
quantities are those presented in Refs.\cite{relic,direct,up}.
The experimental constraints have been updated to include  
the latest LEP2 data\cite{LEP2}. Only the predictions 
for the $\bar p/p$ ratio are not given here, since they 
are presented and discussed in an accompanying paper 
in these Proceedings \cite{mignola}. 
Our results for the neutralino relic abundance and detection rates are 
provided in the form of scatter plots. 
The independent  parameters of the 
model have been varied in the following ranges:

\begin{eqnarray}
10\;\mbox{GeV} &\leq M_2 &\leq  500\;\mbox{GeV}\nonumber\\
10\;\mbox{GeV} &\leq \mu &\leq  500\;\mbox{GeV}\nonumber\\
50\;\mbox{GeV} &\leq m_A &\leq  500\;\mbox{GeV}\nonumber\\
100\;\mbox{GeV} &\leq m_0 &\leq  500\;\mbox{GeV}\nonumber\\ 
100\;\mbox{GeV} &\leq m_{\tilde t}  &\leq  500\;\mbox{GeV}\nonumber\\ 
-3 &\leq A &\leq +3\nonumber\\
1.1 &\leq \tan \beta &\leq 55
\label{eq:range}
\end{eqnarray}

Let us mention here that in more sophisticated supersymmetric models the 
properties at the $M_Z$ scale are derived from properties at the 
Gran Unification scale $M_{GUT}$ by employing the renormalization 
group equations. One of the nicest features of this approach is 
the link between the supersymmetry breaking and the EWSB, 
which is induced radiatively. These supersymmetric  schemes are  
usually based on  assumptions of unification at $M_{GUT}$, not only for 
the gauge 
coupling constants, but also for  Yukawa couplings and for the 
soft-breaking parameters (scalar masses, gaugino masses, trilinear 
couplings).  These assumptions have very strong consequences 
for the 
neutralino phenomenology at the $M_Z$ scale. Relaxation of the universality 
assumptions, for instance the one concerning the scalar masses \cite{op,pol}, 
 may substantially modify the main properties of relic 
neutralinos \cite{ours,our}. We do not discuss these more refined 
supersymmetric models in this paper; only 
at the end of the next section we will 
compare some results obtained in the MSSM with the ones obtainable in 
 theoretical schemes with and without universality conditions
 on the soft scalar masses \cite{gaugino}.

\section{Relic abundance and detection rates in MSSM} 
   Now we present  our main results. Fig.\ref{fig2} shows the 
neutralino relic abundance $\Omega_{\chi} h^2$ as a function 
of $m_{\chi}$ in the form of a scatter plot 
obtained by varying the 
parameters  over the ranges of Eq.(\ref{eq:range}); 
$\Omega_{\chi} h^2$ has been calculated with the method illustrated in 
Ref.\cite{relic}. 
 
\begin{figure}[htb]
\epsfxsize=10cm
\vspace{14cm}
\caption{Neutralino relic abundance as a function of the neutralino 
mass. The solid (dashed) line delimits the region where neutralino
configurations with $\tan \beta$ = 55 and $m_A$ = 50-60  GeV (with 
$\tan \beta$ = 1.1 and $m_A$ = 500 GeV) are located.
\label{fig2}}
\end{figure}
  
The configurations giving an $\Omega_{\chi} h^2$ in conflict with 
the cosmological bound have already been removed. As expected, 
large values of the relic abundance are provided by configurations with 
 small $\tan \beta$ and large  $m_A$, since under 
these circumstances the 
neutralino pair-annihilation cross section is small. 
For the same reason large $\tan \beta$ and small  $m_A$ 
entail small values of $\Omega_{\chi} h^2$. 
By way of example, in Fig.\ref{fig2} two contour lines denote the values of 
$\Omega_{\chi} h^2$ for two subsets of neutralino configurations 
for extreme values of $\tan \beta$ and $m_A$. 

We turn now to the presentation of detection rates for two classes 
of searches for relic neutralinos: 

\begin{itemize}
\item Direct detection, i.e. measurement of the energy released by 
a neutralino in its scattering off a nucleus in an appropriate detector, 
by using very different experimental techniques \cite{jkg}. Some of the most
recent experimental results are given in 
Refs.\cite{heidelberg,boulby,rome,milano,zaragoza,sarsa}.
The rates for direct detection are calculated here as explained in 
Ref.\cite{direct}. 

\item  Detection of the possible signals consisting  of up-going muons 
through a neutrino telescope generated by neutrinos produced by pair
annihilations of neutralinos captured and accumulated inside the Earth and the
Sun. The evaluation of the muon fluxes, which is a rather elaborate multistep 
process, has been performed here according to the procedure described in Ref.
\cite{up}, to which we refer for details of the calculations and for 
 references. 

\end{itemize}

For a presentation of the results concerning a possible excess in 
the $\bar p/p$ ratio  in the cosmic rays we refer to Ref.\cite{mignola}. 

We remind that the detection  rates for both direct detection and 
detection of the up-going muon fluxes are proportional to the neutralino 
local (solar neighborhood) density $\rho_{\chi}$. To assign a value 
to this quantity we employ  here the following rescaling 
recipe \cite{gaisser}. For each point of the parameter
space, we take into account the relevant value of the cosmological neutralino
relic density. When $\Omega_\chi h^2$ is larger than a minimal
$(\Omega h^2)_{min}$, compatible with observational data and with large-scale 
structure calculations, we simply put $\rho_\chi=\rho_l$.
When $\Omega_\chi h^2$ turns out  to be less than $(\Omega h^2)_{min}$, 
and then the neutralino may only provide a fractional contribution
${\Omega_\chi h^2 / (\Omega h^2)_{min}} \equiv  \xi$
 to $\Omega h^2$, we take $\rho_\chi = \rho_l \xi$.
The value to be assigned to $(\Omega h^2)_{min}$ is
somewhat arbitrary, in the range 
$0.03 \lsim (\Omega h^2)_{min} \lsim 0.3$. In the present paper we have used 
$(\Omega h^2)_{min} = 0.03$. As far as the value of $\rho_l$ is concerned, we
have taken the  representative value $\rho_l = 0.5~{\rm GeV \cdot cm^{-3}}$. 
This corresponds to the
central value of a recent determination of $\rho_l$, based on 
a flattened dark matter distribution and microlensing data\cite{turner}: 
$\rho_l = 0.51_{-0.17}^{+0.21}~{\rm GeV \cdot cm^{-3}}$.

Fig.\ref{fig3} shows a comparison between 
one of the most stringent 
experimental  upper bounds: 
$R_{Ge}^{expt} (6\; \mbox{KeV} \leq E_{ee} \leq 8 \; \mbox{KeV})$ = 
1.2 counts/(Kg day)
(this is extracted from the experimental data of Ref. \cite{zaragoza}) and the 
 predicted values for the same quantity. 

\begin{figure}[htb]
\epsfxsize=10cm
\vspace{14cm}
\caption{ $R_{Ge}$ (as defined in the text) versus $m_{\chi}$. 
The contour lines have the same meaning as in Fig.\ref{fig2}. \label{fig3}}
\end{figure}
  
These results are also reported in Fig.\ref{fig4} in a plot versus 
$\Omega_{\chi} h^2$, 
 to show the expected correlation between the signal and the 
relic abundance.

\begin{figure}[htb]
\epsfxsize=10cm
\vspace{14cm}
\caption{$R_{Ge}$ versus the relic abundance. \label{fig4}}
\end{figure}

In Fig.\ref{fig3} 
we notice that for some configurations, with large $\tan \beta$ and 
small $m_A$, the predicted rates are above the experimental limit. 
However, we emphasize that this feature cannot  be used to 
exclude these neutralino configurations, since one has to take into account 
the large 
uncertainties affecting  some of the astrophysical (and cosmological) 
quantities which enter in the evaluation of the detection rates. 
Table 1 reports two sets of 
values for the relevant quantities of this kind. 
Set I corresponds essentially to the central values for the
various parameters, whereas set II corresponds to those values of the
parameters, which, within the 
relevant allowed ranges,  provide the lowest detection rates 
(once the supersymmetric variables are fixed).

\begin{table}
\caption{Values of the astrophysical and cosmological parameters 
relevant to direct detection rates. 
$V_{r.m.s.}$ denotes the root mean square velocity of the neutralino Maxwellian
velocity distribution in the halo; $V_{esc}$ is the neutralino escape velocity
and $V_{\odot}$ is the velocity of the Sun around the galactic centre; 
$\rho_{loc}$ denotes the local dark matter density and $(\Omega h^2)_{min}$ the
minimal value of $\Omega h^2$. The values of set I are the median values of
the various parameters, the values of set II are the extreme values of the
parameters which, within the physical ranges, provide the lowest estimates of 
the direct rates
(once the supersymmetric parameters are fixed).
}
\begin{center}
\begin{tabular}{|c|c|c|}   \hline
 &  Set I &  Set II \\ \hline
$V_{r.m.s}(\rm km \cdot s^{-1}$) & 270 & 245 \\ \hline
$V_{esc}(\rm km \cdot s^{-1}$)   & 650 & 450 \\ \hline
$V_{\odot}(\rm km \cdot s^{-1}$) & 232 & 212 \\ \hline
$\rho_{loc}(\rm GeV \cdot cm^{-3}$) & 0.5 & 0.2 \\ \hline
$(\Omega h^2)_{min}$            & 0.03 & 0.3 \\ \hline
\end{tabular}
\end{center}
\end{table}

For all the results of the present review, unless 
specified differently, the set I has been employed. This is, for instance, 
 the case 
for the detection rates reported in Fig.\ref{fig3}.
 Now it turns out that, if the 
calculation of the rate 
$R_{Ge}^{expt}$ (6 KeV $\leq E_{ee} \leq$ 8 KeV) is performed using set II, 
all predicted rates are below the experimental upper bound. However, it 
has to be stressed that the maximum value of the rate falls short of the 
limit only by a factor of 2. 
This obviously implies that even a moderate improvement in experimental 
 sensitivities in direct searches may provide an essential information about 
the neutralino parameter space. 

A more direct illustration of this point is provided by an inspection 
of the neutralino-nucleon cross section.  Fig.\ref{fig5}  gives a comparison of
the predicted values for $\xi \cdot \sigma_{scalar}^{(n)}$ with the relevant
upper bound (as derived from the current experimental data 
Refs.\cite{heidelberg,boulby,rome,milano,zaragoza}). 
 
\begin{figure}[htb]
\epsfxsize=10cm
\vspace{6cm}
\caption{The neutralino-nucleon scalar cross section 
multiplied by the scaling factor $\xi$ versus 
$m_{\chi}$. Set I of Table 1 is used. 
The contour lines have the same meaning as in Fig.\ref{fig2}. \label{fig5}}
\end{figure}
  
$\sigma_{scalar}^{(n)}$
stands for the neutralino-nucleon scalar cross-section, responsible for
coherent effects in processes involving nuclei. Here set I has been used. 
The results of the same calculation with set II for the astrophysical and 
cosmological parameters are shown in Fig.\ref{fig6}. 
\begin{figure}[htb]
\epsfxsize=10cm
\vspace{6cm}
\caption{Same as in Fig.\ref{fig5}, except that here set II of Table 1 
is used. \label{fig6}}
\end{figure}
We notice that no predicted values
are above the experimental upper bound;  however, some neutralino configurations
are close to it. We do not consider here explicit upper bounds on the 
spin-dependent neutralino-nucleon cross section as obtainable from present
experimental data. 
This point will be  discussed in detail in a forthcoming paper \cite{bds}. 

Let us turn now to the indirect measurements, based on the  outcome of 
high-energy neutrinos from neutralino pair annihilation in celestial bodies. 
Here we limit ourselves to the emission from the center of the Earth. 
The  flux of the ensuing up-going muons, 
integrated over muon energies above 
1 GeV and over a cone of half aperture of $30^{\circ}$ centered at 
the nadir,  is shown in Fig.\ref{fig7}.  
\begin{figure}[htb]
\epsfxsize=10cm
\vspace{14cm}
\caption{$\Phi_{\mu}^{Earth}$ versus $m_{\chi}$. 
The contour lines have the same meaning as in Fig.\ref{fig2}. \label{fig7}}
\end{figure} 
The horizontal line denotes 
the Baksan upper limit: 
$\Phi_{\mu}^{Earth} \leq 2.1 \times 10^{-14} {\rm cm^{-2} s^{-1}} 
(90 \% C.L.)$ \cite{bbb}. 

Also in this case, many configurations provide fluxes in excess of 
the experimental limit. However, if set II of Table 1 is employed, then the 
number of neutralino configurations staying above the experimental bound 
turns out to be 
marginal. It then follows  that present data 
from neutrino telescopes are not yet
capable of excluding physical regions of the supersymmetric parameter space 
(contrary to the conclusions of Ref.\cite{bbb}). 

Fig.\ref{fig8} displays $\Phi_{\mu}^{Earth}$ 
versus $R_{Ge}$ to show the expected 
correlation between these two quantities. Also displayed are 
 the
locations of configurations which would be selected by a 
 model with 
 universality assumptions at $M_{GUT}$ and by one where the unification 
condition on 
soft scalar masses has been somewhat relaxed \cite{ours}.
This illustrates 
how strong are the constraints imposed by strict universality conditions 
at $M_{GUT}$ and how sensitive are the signals to these assumptions.

\begin{figure}[htb]
\epsfxsize=10cm
\vspace{14cm}
\caption{$\Phi_{\mu}^{Earth}$ versus $R_{Ge}$. 
The solid line delimits the location of configurations for  the MSSM 
considered in this paper. The diamonds (the dots) denote the 
values for a 
model with strict universality assumptions at $M_{GUT}$ (with a deviation from 
unification condition on the soft scalar masses).  \label{fig8}}
\end{figure}

From what we have discussed in this part, we can conclude that the
present sensitivities for the experimental searches examined above are not yet
in the position of constraining the supersymmetric model. However, the 
foreseable improvements in these sensitivities will make soon possible 
an  exploration of some regions of the neutralino parameter space. 

\section{A Higgsino-like relic neutralino}

It has recently been conjectured \cite{ambro1} that 
 the CDF event
$p \bar p \rightarrow e^+ e^- \gamma \gamma \rlap{/} {\rm E_T}$ \cite{cdf} 
is due to a decay chain involving 
two  neutralino states (the lightest and the next-to-lightest ones). 
The lightest neutralino (that we denote as $\chi_{AKM}$) has been further 
considered in Ref.\cite{kane} 
as a candidate for cold dark matter. 
If the neutralino interpretation of Ref.\cite{ambro1} is correct, then 
the neutralino would be confined in a rather narrow region of the
supersymmetric parameter space, at least as far as some parameters (such as 
$M_2, \mu, \tan \beta$) are concerned. An important question is whether or not
a relic  $\chi_{AKM}$ could be detected either directly or indirectly. 
Here we only
report a few results of a thorough investigation carried out in 
Ref.\cite{ol}, to which we refer for further details 
(some detection signals for 
$\chi_{AKM}$ have also been investigated in Refs.\cite{kane,fk} for a limited
domain of the $\chi_{AKM}$ parameter space and under the hypothesis that 
$\chi_{AKM}$ provides a large contribution to $\Omega$). 

In Ref.\cite{ol} it is shown that 
for most regions of the parameter space the detectability of a relic 
$\chi_{AKM}$ would require   quite substantial improvements in current 
experimental sensitivities. 
Some favorable perspectives for investigating a region of 
the $\chi_{AKM}$ parameter space, around the maximal $\tan \beta$ value allowed
by the model of Ref.\cite{ambro1} 
({\it i.e.}, $\tan \beta \simeq 2.5$), 
are offered by  measurements 
of neutrino fluxes from the 
center of the Earth and of an excess of $\bar p/p$ in cosmic rays. To
illustrate this point we display in Fig.\ref{fig9} 
$\bar p/p$ versus $\Phi_{\mu}^{Earth}$
for $\chi_{AKM}$ 
configurations with 1.5 $\leq \tan \beta \leq$ 2.5. 

\begin{figure}[htb]
\epsfxsize=10cm
\vspace{14cm}
\caption{$\bar p/p$ versus $\Phi_{\mu}^{Earth}$ for $\chi_{AKM}$ configurations.
The meaning of the symbols is explained in the text. \label{fig9}}
\end{figure}
  
Diamonds and crosses denote the values of the  signals when these are 
within two orders of magnitude from the current value of at least one of the 
relevant experimental upper bounds (at 90 \% C.L.). 
Diamonds (crosses) denote values obtained by 
evaluating 
the neutralino relic abundance with the exact expression (with a low-velocity
approximation). 
Dots denote the values of the signals for the other
configurations, calculated in the low-velocity approximation only.  
The experimental bounds  are displayed by the 
 horizontal and the vertical lines:  
$\bar p/p \leq 7.5 \times 10^{-5}$ \cite{mit}, 
$\Phi_{\mu}^{Earth} \leq 2.1 \times 10^{-14} {\rm cm^{-2} s^{-1}} 
(90 \% C.L.)$ \cite{bbb}. 

Let us finally remark that the conjecture of Ref.\cite{ambro1} 
is certainly very intriguing; if correct, it would have important 
consequences 
in particle physics and possibly also in cosmology. However,   great caution is 
in order, because of the existence of only a single event of the type  
$p \bar p \rightarrow e^+ e^- \gamma \gamma \rlap{/} {\rm E_T}$ 
and of the
non-uniqueness in its interpretation \cite{ambro1,dimo,dtw,ln,hty,moha}.

{\bf Acknowledgements }

I wish to express my warmest thanks to Fiorenza Donato, Giulio Mignola, 
Nicolao Fornengo and Stefano Scopel
for their most valuable collaboration in the preparation of this report. 
This work was 
supported in part by the Research Funds of the Ministero dell'Universit\`a 
e della Ricerca Scientifica e Tecnologica.

\vfill
\eject



\begin{thebibliography}{99}


\bibitem{ours} V. Berezinsky, A. Bottino, J. Ellis, N. Fornengo,
G. Mignola and S. Scopel, 
{\it Astroparticle Physics} {\bf 5} (1996) 333. 

\bibitem{Susy} H. P. Nilles, {\it Phys. Rep.} {\bf 110} (1984) 1; \hfill \break
H. E. Haber and G. L. Kane, {\it Phys. Rep.} {\bf 117} (1985) 75; \hfill \break
R. Barbieri {\it Riv. Nuovo Cim.} {\bf 11} (1988) 1.


\bibitem{oneloop} S. Coleman and E. Weinberg, {\it Phys. Rev.} {\bf D7} (1973)
1888; \hfill \break
G. Gamberini, G. Ridolfi and F. Zwirner, {\it Nucl. Phys.} {\bf B331} (1990)
331; \hfill \break
R. Arnowitt and P. Nath, {\it Phys. Rev.}
{\bf D46} (1992) 3981.



\bibitem{LEP2} OPAL Collaboration, CERN Preprint CERN-PPE/96 October 4, 1996; 
M. Martinez, W. De Boer, M. Pohl, A.T. Watson representing the ALEPH, 
DELPHI, L3 and OPAL 
Collaborations, joint seminar at CERN, October 8, 1996. 

\bibitem{alam} M.S. Alam et al. (CLEO coll.), {\it Phys. Rev. Lett.} 
{\bf 74} (1995) 2885. 




\bibitem{jkg} G. Jungman, M. Kamionkowski and K. Griest, {\em Phys. Rep.} 
{\bf 267} (1996) 195
and references quoted therein. 

\bibitem{bg} L. Bergstr\"om and P. Gondolo, {\it Astroparticle Physics} 
{\bf 5} (1966) 263.


\bibitem{all} The  results  presented in this paper 
have been obtained 
in collaboration with F. Donato, N. Fornengo, G. Mignola 
and S. Scopel. 




\bibitem{relic} A. Bottino, V. de Alfaro, 
N. Fornengo, G. Mignola and M. Pignone,
{\em Astroparticle Physics} {\bf 2} (1994)  67.

\bibitem{direct} A. Bottino, V. de Alfaro, N. Fornengo, G. Mignola and 
S. Scopel, 
{\em Astroparticle Physics} {\bf 2} (1994)  77.

\bibitem{up} A. Bottino, N. Fornengo, G. Mignola and L. Moscoso, 
{\em Astroparticle Physics} {\bf 3} (1995) 65.



\bibitem{mignola} G. Mignola: {\it Cosmic antiprotons from neutralino 
annihilation in the galactic halo}, these Proceedings. 


\bibitem{op} M. Olechowski and S.Pokorski, {\it Phys. Lett.} {\bf B344}
(1995) 201.


\bibitem{pol}
N. Polonsky and A. Pomarol, {\it Phys. Rev. Lett.} {\bf 73} (1994) 2292;
{\it Phys. Rev.} {\bf D51} (1995) 6532; \hfill \break
D. Matalliotakis and  H. P. Nilles, {\it Nucl. Phys.} {\bf B435}
(1995) 115; \hfill\break
A. Pomarol and S. Dimopoulos, {\it Nucl. Phys.} {\bf 453} (1995) 83;
\hfill\break
H. Murayama, Berkeley preprint LBL-36962, hep-ph/9503392.

\bibitem{our} V. Berezinsky,  A. Bottino, J. Ellis, N. Fornengo,
G. Mignola and S. Scopel, {\it Astroparticle Physics}  {\bf 5} (1996) 1.


\bibitem{gaugino} For the effects on the neutralino properties of a 
relaxation in the unification assumption for the gaugino masses see, 
for instance: K. Griest and L. Roskowski, {\it Phys. Rev.} {\bf D46} 
(1992) 3309 and V.A. Bednyakov, S.G. Kovalenko and H.V. Klapdor-Kleingrothaus, 
 hep-ph/9608241.


\bibitem{heidelberg} M. Beck et al. (Heidelberg-Moscow), {\em Phys. Lett.}
{\bf B336} (1994) 141.

\bibitem{boulby} P.F. Smith et al., {\em Phys. Lett.} {\bf B379} (1996) 299.


\bibitem{rome} P. Belli et al., {\em Nucl. Phys.} {\bf B} 
(Proc. Suppl.) {\bf 48} (1996) 60; R. Bernabei, these Proceedings. 

\bibitem{milano} A. Alessandrello et al., {\em Phys. Lett.} {\bf B384} 
(1996) 316.

\bibitem{zaragoza} A.K. Drukier et al., {\em Nucl. Phys.} {\bf A} 
(Proc. Suppl.) {\bf 28} (1992) 293.

\bibitem{sarsa} M.L. Sarsa et al., {\it Nucl. Phys. (Proc. Suppl.)} 
{\bf B48} (1996) 73

\bibitem{gaisser} T.K. Gaisser, G. Steigman and S. Tilav, {\em Phys. Rev.}
 {\bf D34} (1986) 2206.

\bibitem{turner} E.I. Gates, G. Gyuk and M. Turner, {\em Ap.J.} 
 {\bf 449} (1995) L123.


\bibitem{bds} A.Bottino, F. Donato, G. Mignola and S. Scopel, in preparation.


\bibitem{bbb} M.M. Boliev at al. (Baksan Underground Telescope), 
{\it Nucl. Phys.} 
{\bf B} (Proc. Suppl.) {\bf 48} (1996) 83.





\bibitem{ambro1} S. Ambrosanio, G.L. Kane, G.D. Kribs, S. Martin and S. Mrenna,
 {\em Phys. Rev. Lett.} {\bf 76} (1996) 3498 and hep-ph/9607414. 


\bibitem{cdf} See, for instance, L. Nodulman (CDF Collaboration), talk at the
International Europhysics Conference on High Energy Physics, Brussels, July
1996.



\bibitem{kane} G.L. Kane and J.D. Wells, {\em Phys. Rev. Lett.} 
{\bf 76} (1996) 4458.

\bibitem{ol} A. Bottino, N. Fornengo, G. Mignola, M. Olechowski and S. Scopel, 
astro-ph/9611030.


\bibitem{fk} K. Freese and M. Kamionkowski, hep-ph/9609370.

\bibitem{mit} J.W. Mitchell et al., {\em Phys. Rev. Lett.} {\bf 76} (1996)
3057.


\bibitem{dimo} S. Dimopoulos, M. Dine, S. Raby and S. Thomas, {\em 
Phys. Rev. Lett.} 
{\bf 76} (1996) 3494.




\bibitem{dtw} S. Dimopoulos, S. Thomas and J.D. Wells, {\em Phys. Rev.} {\bf
D54} (1996) 3283.

\bibitem{ln} J.L. Lopez and D.V. Nanopoulos, hep-ph/9607220.

\bibitem{hty} J. Hisano, K. Tobe and T. Yanagida, hep-ph/9607234.


\bibitem{moha} G. Bhattacharyya and R. Mohapatra, {\em Phys. Rev.} {\bf
D54} (1996) 4204.


\end{thebibliography}
\end{document}